\documentclass[preprintnumbers,11pt,prd,showpacs]{revtex4}
\usepackage{graphicx}
\usepackage{amssymb}
\usepackage{epstopdf}
\usepackage{amsmath}
\usepackage{bm}
\begin{document}
\preprint{BU-HEPP}

\title{Polynomial Subtraction Method for Disconnected Quark Loops}\thanks{This work is partially supported by a grant from Baylor University Research Committee.}

\author{Quan Liu}
\thanks{quan\_liu@baylor.edu, Baylor Physics Department}
\author{Walter Wilcox}
\thanks{walter\_wilcox@baylor.edu, Baylor Physics Department}
\author{Ron Morgan}
\thanks{ronald\_morgan@baylor.edu, Baylor Mathematics Department}
\affiliation{Department of Physics, Baylor University, Waco, TX 76798-7316}
\affiliation{Department of Mathematics, Baylor University, Waco, TX 76798-7316}
\begin{abstract}
The polynomial subtraction method, a new numerical approach for reducing the noise variance of Lattice QCD disconnected matrix elements calculation, is introduced in this paper. We use the MinRes polynomial expansion of the QCD matrix as the approximation to the matrix inverse and get a significant reduction in the variance calculation. We compare our results with that of the perturbative subtraction and find that the new strategy yields a faster decrease in variance which increases with quark mass. 
\end{abstract}


\maketitle
\newpage

\section{Introduction}\label{Sone}
Many Lattice QCD calculations require the  evaluation of quark matrix elements of disconnected loops. Examples include the prototype calculation of the disconnected part of the nucleon electromagnetic form factors \cite{wilcoxtm}, the strangeness and charmness contents of the nucleon \cite{strange}, and determination of hadronic scattering lengths \cite{scatter}. The exact calculation of the quark matrix elements at each lattice point is extremely difficult and unrealistic with current computer resources. An alternative approach is to calculate the unbiased stochastic estimates \cite{noisestat1,noisestat2,wwnoise,Bernardson1993he} of the operator. This method utilizes noise theory, which based upon the projection of the matrix elements using random noise input. The prevailing numerical methods include eigenvalue subtraction method\cite{vicpaper} and perturbative subtraction method\cite{Wilcoxpert}. In this paper, we will present a new approach we term the polynomial subtraction method. We will start by a brief review of the noise theory in section \ref{Stwo}. The idea of the subtraction methods are introduced in section \ref{Sthree}. The correction strategy is discussed in section \ref{Sfour}. In section \ref{Sfive}, we present the numerical test results for the polynomial subtraction method. It is shown that this new method outperforms the traditional perturbative subtraction method consistently for small to medium $\kappa$ values with minimal extra computational expenses. A conclusion is made in section \ref{Ssix} based on our numerical tests.
\section{Noise Theory}\label{Stwo}
Before we talk about the subtraction methods, let us briefly review noise theory. Consider a system which can be described as
\begin{equation}
Mx=\eta,
\end{equation}
where $M$ is the $N \times N$ quark matrix, $x$ is the solution vector and  $\eta$  is a random noise vector used to project the matrix elements, with
\begin{equation}
\langle \eta_{i} \rangle=0,\langle \eta_{i}\eta_{j} \rangle=\delta_{ij},
\end{equation}
where an averaging is over all the noises is used. The matrix element, $M^{-1}_{ij}$, can be calculate from
\begin{equation}
\langle \eta_{j}x_{i} \rangle=\sum_{k}M^{-1}_{ik}\langle \eta_{j}\eta_{k}\rangle=M^{-1}_{ij}.
\end{equation}
Now we want to evaluate the variance of this method. The quantity we are most interested in is the trace, so we will focus on the variance of this quantity. Define
\begin{equation}
X_{mn}\equiv \frac{1}{L}\sum^{L}_{l=1}\eta^{(l)}_{m} \eta^{(l)*}_{n}
\end{equation}
for $(m,n=1,2, \dots, N)$, where $N$ is the dimension of the matrix and $L$ is the number of noise vectors used. We have $X_{mn}=X^{*}_{nm}$ and $\langle X_{mn}\rangle=\delta_{mn}$. It can be shown \cite{noisestat3} that
\begin{eqnarray}
V[Tr\{QX\}] &\equiv& \langle |\sum_{m,n} q_{mn}X_{mn}-Tr\{Q\}|^{2}\rangle \nonumber \\
                   &=&\sum_{n}\langle |X_{nn}-1|^{2}\rangle \langle |q_{nn}|\rangle^{2} \nonumber \\
                   & &+ \sum_{m \ne n}\left(\langle |X_{mn}|^{2}\rangle |q_{mn}^{2}|+\langle (X_{mn}^{2})\rangle q_{mn}q^{*}_{nm}\right) , \label{variance}
\end{eqnarray}
where Q is the matrix-representation of an operator. First, let's consider a general real noise. The constraints are:
\begin{equation}
\langle |X_{mn}|^{2}\rangle = \langle (X_{mn})^{2}\rangle =\frac{1}{L} \label{constaints}
\end{equation}
for $m\ne n$. Using Eq.(\ref{variance}), the variance for general real noise is:
\begin{equation}
\begin{split}
V[Tr\{QX_{real}\}]=& \frac{1}{L}\sum_{m\ne n}(|q_{mn}|^{2}+q_{mn}q^{*}_{nm}) \\
          &+ \sum_{n} \langle |X_{nn}-1|^{2}\rangle |q_{nn}|^{2}.
\end{split}
\end{equation}
The $Z(2)$ noise also has Eq.(\ref{constaints}), for $m\ne n$ and an extra constraint $\langle |X_{nn}-1|^{2}\rangle =0$. The result for $Z(2)$ noise is:
\begin{equation}
V[Tr{QX_{Z(2)}}]=\frac{1}{L}\sum_{m\ne n}(|q_{mn}|^{2}+q_{mn}q^{*}_{nm}).
\end{equation}
For the $Z(N)(N\ge 3)$ noise, the constraints become:
\begin{equation}
\langle |X_{mn}|^{2}\rangle =\frac{1}{L},\langle (X_{mn})^{2}\rangle =0,\langle |X_{nn}-1|^{2}\rangle =0 \label{constaints2}.
\end{equation}
Thus the variance is:
\begin{equation}
V[Tr\{QX_{Z(N)}\}]=\frac{1}{L}\sum_{m\ne n}|q_{mn}|^{2}.
\end{equation}
Generally speaking, there's no fixed relationship between $Z(2)$ and $Z(N)$.In this paper, however, we assume the phases of $q_{mn}$ and $q^{*}_{nm}$ are uncorrelated. Then we have $V[Tr\{QX_{Z(2)}\}] \approx V[Tr{QX_{Z(N)}}]$, $(N \ge 3)$. So we can conclude that the variance of the trace calculation is proportional to the sum of the off-diagnal elements of the quark matrix. In this paper, all the work is done with the $Z(4)$ noise. The idea of subtraction method is to find a traceless matrix which has similar off-diagnal elements as the matrix we want to calculate. Consider matrix $\tilde{Q}$ such that
\begin{equation}
\langle Tr\{ \tilde{Q} \}\rangle=0.
\end{equation}
Thus, $\langle Tr \{(Q-\tilde{Q}) X\}\rangle =\langle Tr\{Q\}\rangle$, for traceless $\tilde{Q}$. If the off-diagnal elements in $\tilde{Q}$ are close to those of $Q$, the variance will thus be reduced.
\section{Subtraction Method}\label{Sthree}
The matrix we need to calculate is given by 
\begin{equation}
(M^{-1})_{I J}=\frac{1}{\delta_{I J}-\kappa P_{I J}},
\end{equation}
where $\{I J\}$ are collective indices and
\begin{equation}
P_{I J}=\sum_{\mu}[(1-\gamma_{\mu})U_{\mu}(x)\delta_{x,y-a_{\mu}}+(1+\gamma_{\mu})U^{\dagger}_{\mu}(x-a_{\mu} \delta_{x,y+a_{\mu}})].
\end{equation}
In general, the expectation value of an operator is given as
\begin{eqnarray}
\langle \bar{\psi} O \psi \rangle &=&-Tr(OM^{-1}) \nonumber \\
                                &=&-\sum_{i}O \langle x_{i} \eta_{i} \rangle \nonumber \\
                                &=&-\sum_{i}\frac{1}{L}\sum^{L}_{j}Ox^{j}_{i}\eta^{*j}_{i} \nonumber \\
                                &=&-\sum_{i}\frac{1}{L}\sum^{L}_{j} O \left( \sum_{k}M^{-1}_{ik}\eta^{j}_{k} \right) \eta^{*j}_{i} \label{expV}
\end{eqnarray}
The idea is to find an appximation, $\tilde{M}^{-1}$, whose off-diagnal elements mimic the ones in $M^{-1}$. We can insert the $\tilde{M}^{-1}$ into Eq.(\ref{expV}) and get:
\begin{eqnarray}
\langle \bar{\psi} O \psi \rangle &=&-\sum_{i}\frac{1}{L}\sum^{L}_{j} \eta^{*j}_{i}\sum_{k} O\left( M^{-1}_{ik}-\tilde{M}^{-1}_{ik}\right) \eta^{j}_{k}-Tr(O\tilde{M}^{-1}) \nonumber \\
&=&-\frac{1}{L}\sum^{L}_{j}\left( \eta^{j}\centerdot O \left(x^{j}-\tilde{M}^{-1}\eta^{j} \right) \right)-Tr(O\tilde{M}^{-1}) \label{expV1}
\end{eqnarray}
Note that in the second step, I change the notation to the dot product form. As discussed in the first section,the introduction of $\tilde{M}^{-1}$ will decrease the variance of the calculation. But the problem is: the $\tilde{M}^{-1}$ is not traceless in most cases. As shown in Eq.(\ref{expV1}), we have to subtract $Tr(O\tilde{M}^{-1})$ to get the  unbiased expectation value.
Some strategies are developed to build different kinds of $\tilde{M}^{-1}$, such as perturbative subtraction \cite{Wilcoxpert,deanphd} and eigenspectrum subtraction \cite{vicpaper}. The eigenspectrum subtraction method, though most promising, is now limited to small lattice tests due to technical problems. In this paper, we will introduce a new technique, which is called polynomial subtraction method  and focus on the comparison between the perturbative subtraction method and the polynomial subtraction method.

The idea of perturbative method is to expand the $M^{-1}$ in geometric series\cite{Wilcoxpert}:
\begin{equation}
\tilde{M}^{-1}_{pert}=I+\kappa P+\kappa^{2}P^{2}+\kappa^{3}P^{3}+\dots
\end{equation}
There are two benefits of this method. First, the $\tilde{M}^{-1}_{pert}$ is easy to build. Second, $Tr(O\tilde{M}^{-1}_{pert})$ is easy to calculate so that it is convenient for us to correct the subtracted expectation value in Eq.(\ref{expV1}). Inspired by this idea, we construct a new $\tilde{M}^{-1}$ by using the minimal residual Polynomial\cite{Minres}. Consider the system:
\begin{equation}
Mx=\eta
\end{equation} 
We want to minimize $||Mx_{t}-b||_{2}$ in the Krylov subspace spanned by $ b,Mb,M^{2}b,M^{3}b,\dots $,that is,
\begin{equation}
\mathcal{K}_{t}=\{b,Mb,M^{2}b,M^{3}b,\dots \}
\end{equation}
Since $x_{t}\in \mathcal{K}_{t}$,we can express $x_{t}$ as,
\begin{eqnarray}
x_{t}&=&a_{0}b+a_{1}Mb+a_{2}M^{2}b+a_{3}M^{3}b+\dots \nonumber \\
         &=&\left(a_{0}+a_{1}M+a_{2}M^{2}+a_{3}M^{3}+\dots \right)b \nonumber \\
         &=&P(M)b
\end{eqnarray}
where $P(M)$ is a polynomial of M. The norm of the residual can be rewritten as:
\begin{equation}
||(MP(M)-I)b||_{2}
\label{res}
\end{equation} 
We can see that when the residual norm is minimized, we have $P(M)\approx M^{-1}$. Consider an $n^{th}$ order polynomial of M, $ P_{n}(M)=a_{0}+a_{1}M+a_{2}M^{2}+\dots +a_{n}M^{n}$. The coefficients $\bm{a}=\{a_{0},a_{1},\dots ,a_{n}\}$ can be determined by solving a small $(n+1)\times (n+1)$ system:
\begin{equation}
\left[ Mb \quad M^{2}b \quad \dots \quad  M^{n+1}b \right]^{\dagger}\left[ Mb \quad M^{2}b  \quad \dots \quad  M^{n+1}b \right]\bm{a}=\left[ Mb \quad M^{2}b \quad \dots \quad  M^{n+1}b \right]^{\dagger}b
\end{equation}
$P(M)$ is the $\tilde{M}^{-1}_{poly}$ we are going to use in the polynomial subtraction method. 

\section{Correction for the Vacuum Expectation Value}\label{Sfour}
\begin{figure}
 \begin{center}
  \includegraphics[scale=.5]{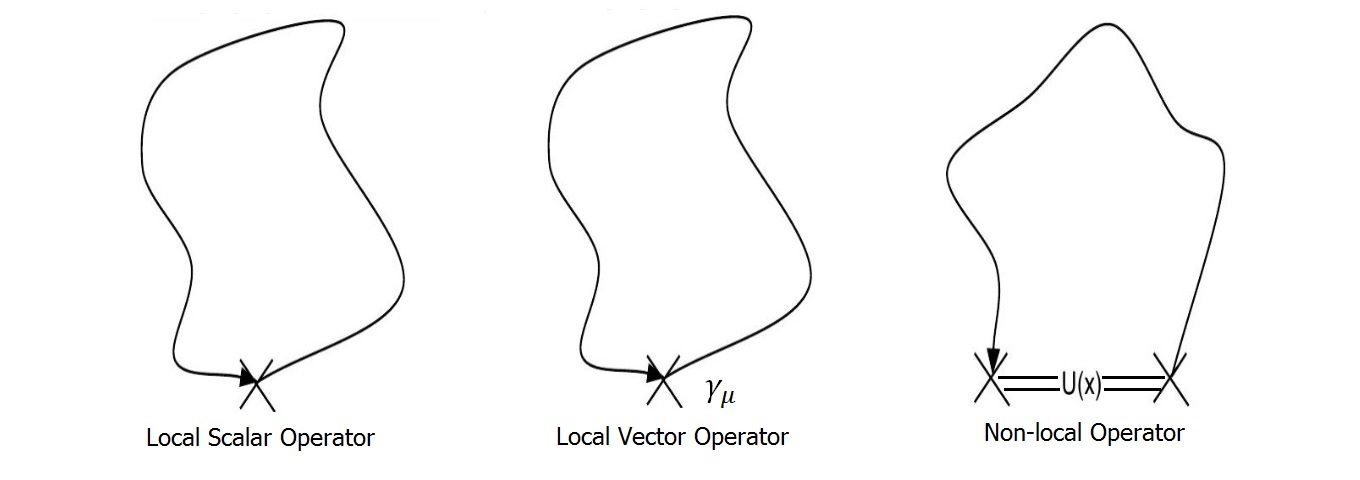}
 \end{center}
 \caption{General diagram of the quark line contributions for local scalar, local vector and non-local operators.}
 \label{fig:general}
\end{figure}

The $\tilde{M}^{-1}s$ involvled in the two subtraction methods are not traceless so the diagnal elements of the matrix $M$ will be changed. This will therefore change the vacuum expectation value and we need to add some correction terms after the subtraction. Notice that only closed loop, gauge invariant objects contribute to the trace in Eq.(\ref{expV1}). In other words, only closed path objects with an area A contribute to the trace in Eq.(\ref{expV1}). The general picture of the local scalar, local vector and non-local operator is given in figure\ref{fig:general}. 

\begin{figure}
 \begin{center}
  \includegraphics[scale=.5]{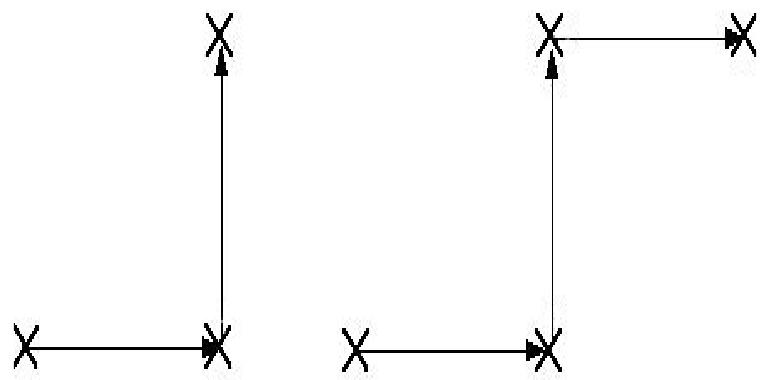}
 \end{center}
 \caption{Perturbative expansion contributions of $O(\kappa^2)$ and $O(\kappa^3)$.}
 \label{fig:23order}
\end{figure}

\begin{figure}
 \begin{center}
  \includegraphics[scale=.5]{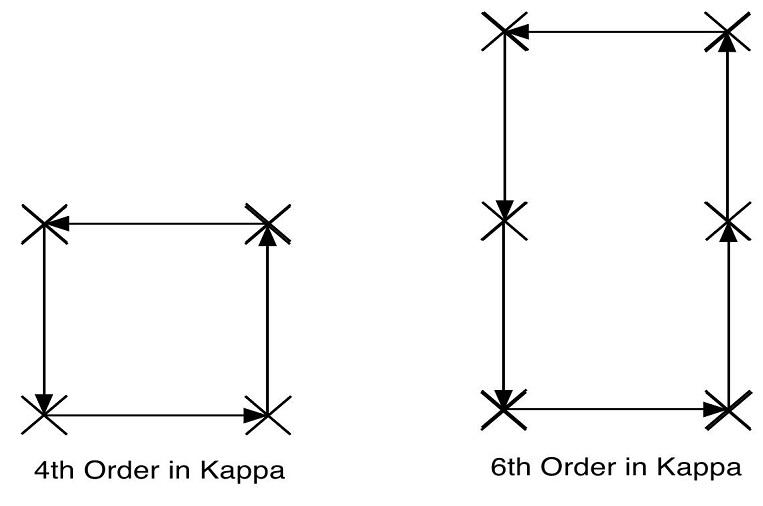}
 \end{center}
 \caption{Perturbative scalar operator contribution at $4^{th}$ and $6^{th}$ order in $\kappa$.}
 \label{4th6thorder}
\end{figure}

\begin{figure}
 \begin{center}
  \includegraphics[scale=.5]{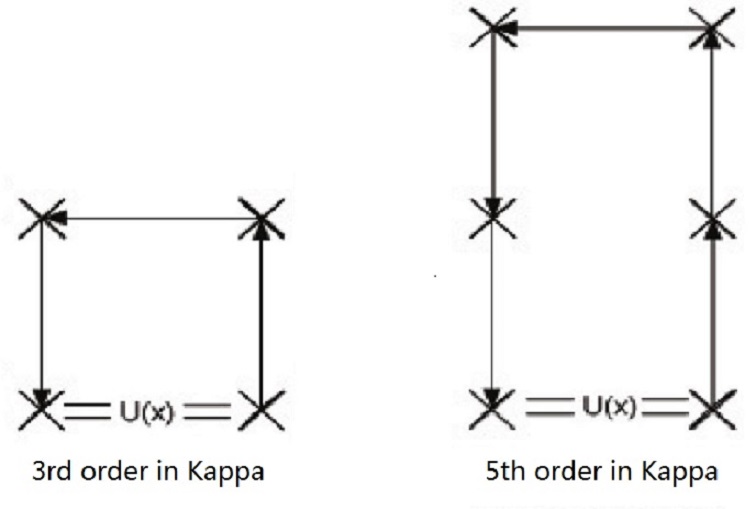}
 \end{center}
 \caption{Perturbative vector operator contribution at $3^{th}$ and $5^{th}$ order in $\kappa$.}
 \label{3rd5thorder}
\end{figure}

The geometric interpretation of the perturbative expansion in figure\ref{fig:23order}\cite{deanphd} shows how each order of $\kappa$ is related to a link. We can easily see from the figure that the local operators require a correction staring at $4^{th}$ order of $\kappa$(the local scalar operators require a correction starting at the $0^{th}$ order of $\kappa$) and non-local operators require a correction starting at $3^{rd}$ order of $\kappa$ because the minimum numer of links required to form closed loops in local operator and non-local operator is 4 and 3 respectively. (There is an implicit order of $\kappa$ in the non-local operator.) Generally speaking, the even orders of $\kappa$ will contribute to the local operators and the odd orders will  contribute to the non-local operators. The examples of closed loops are shown in figure\ref{4th6thorder} and figure\ref{3rd5thorder} .

Although the direct calculation of the $Tr(OM^{-1}_{pert})$ is too expensive, the closed loops can be easily found\cite{vicphd}. We start by solving the system without the noise vector:
\begin{equation}
M_{pert}x=e_{i}
\end{equation}
where $e_{i}$ is the unit vector in the $i^{th}$ direction that spans the space-time-color-Dirac space. And $x$ is found by calculating
\begin{equation}
x=M^{-1}_{pert}e_{i}=(I+\kappa P+\kappa^{2}P^{2}+\kappa^{3}P^{3}+\dots+\kappa^{n}P^{n})e_{i}
\end{equation}
But based on the previous analysis, not all the $O(\kappa ^{n})$ terms contribute to the calculation of the trace for each specific operator. We will explicitly drop the terms that won't contribute to the trace. For example, we will drop the even orders of $\kappa$ when calculating the non-local operators and drop the odd orders of $\kappa$ for the calculation of local scalar and local vector operators. We denote the truncated expansion as $\hat{M}^{-1}_{pert}$. The trace of the correction part is calculated as:
\begin{equation}
Tr(OM^{-1}_{pert})=\sum_{i}\left( e^{*}_{i}O\hat{M}^{-1}_{pert}e_{i}\right)
\end{equation}

The calculation for the correction part of polynomial subtraction is quite similar to that of the perturbative subtraction. Since $M=I+\kappa P$, we can express our $\tilde{M}^{-1}_{poly}$ in terms of $P$:
\begin{eqnarray}
\tilde{M}^{-1}_{poly}&=&a_{0}+a_{1}M+a_{2}M^{2}+a_{3}M^{3}+\dots+a_{n}M^{n} \nonumber \\
                                   &=&a_{0}+a_{1}(I+\kappa P)+a_{2}(I+\kappa P)^{2}+a_{3}(I+\kappa P)^{3}+\dots+a_{n}(I+\kappa P)^{n} \nonumber \\
                                   &=&b_{0}+b_{1}\kappa P+b_{2}\kappa^{2}P^{2}+b_{3}\kappa^{3}P^{3}+\dots+b_{n}\kappa^{n}P^{n}
\end{eqnarray}
The pattern is quite similar to the $\tilde{M}^{-1}_{pert}$ except that the coefficients for $O(\kappa^{n})$ are not ones. These coefficients $\{b_{0},b_{1},\dots,b_{n}\}$ can be easily determined as long as we get the  $\{a_{0},a_{1},\dots,a_{n}\}$.
So we can build the truncated $\hat{M}^{-1}_{poly}$ in the same way as we did for the $\hat{M}^{-1}_{pert}$ by dropping the terms that don't contribute to trace calculation. And the correction term is:
\begin{equation}
Tr(OM^{-1}_{poly})=\sum_{i}\left( e^{*}_{i}O\hat{M}^{-1}_{poly}e_{i}\right).
\end{equation} 
\section{Numerical Result}\label{Sfive}
\begin{figure}
 \begin{center}
  \includegraphics[scale=.75]{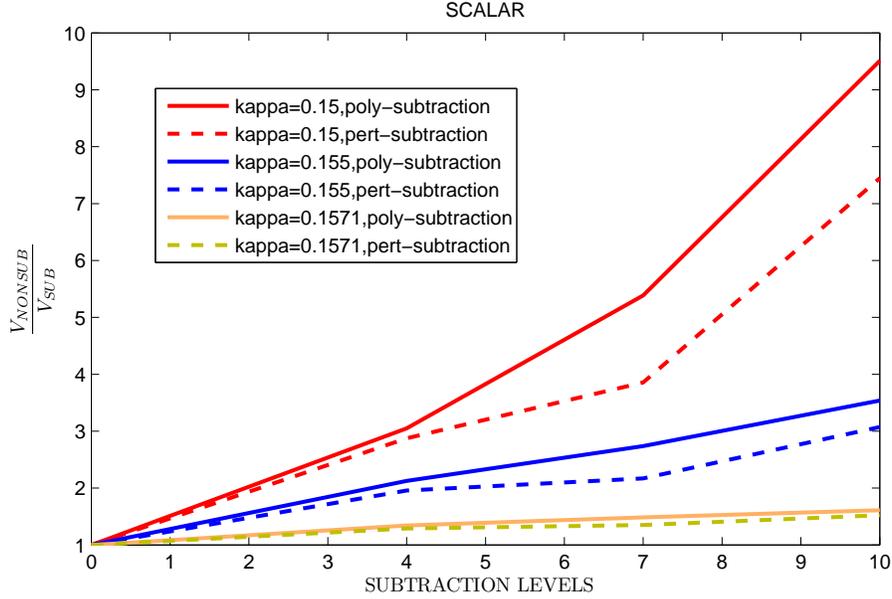}
 \end{center}
 \caption{Subtracted result for local scalar operator.}
 \label{scalar}
\end{figure}

The calculation of the operators are delicate and susceptible to signal degradation from noise by varying degrees, depending on the operator. Each operator is calculated with a real and imaginary part. But due to the quark propagator identity $S=\gamma_{5}S^{\dagger}\gamma_{5}$, only the real or imaginary part, should be non-zero. The identities are (at each lattice site):
\begin{equation}
\label{operators}
 \begin{split}
  \text{Scalar : } &Re\left[ \bar{\psi}(x) \psi(x) \right] \\
  \text{Local Vector : } &Im\left[ \bar{\psi}(x) \gamma_\mu \psi(x) \right] \\
  \text{Pseudoscalar : } &Re\left[ \bar{\psi}(x) \gamma_5 \psi(x) \right] \\
  \text{Axial : } &Re\left[ \bar{\psi}(x) \gamma_5\gamma_\mu \psi(x) \right] \\
  \text{Point-Split Vector : } &\kappa Im\left[ \bar{\psi}(x+a_\mu)(1+\gamma_\mu)U^\dagger_\mu(x) \psi(x)-\bar{\psi}(x)(1-\gamma_\mu)U_\mu(x)\psi(x+a_\mu) \right]
 \end{split}
\end{equation}
In this paper, we will focus on the calculation of local scalar, local vector and point-split vector identities. The work is done by using the quenched Wilson gauge at $\beta=6.0$. We first investigate the effectiveness of the polynomial-subtraction method on a $16^{4}$ lattice with $\kappa=0.15, 0.155, 0.1571$ respectively. Twenty $Z(4)$ noises are employed for each configuration and the results are averaged over 10 configurations. Since we are only interested in the statistical error of a technique, the variance associated between different gauge configurations will be ignored. And the calculation of the exact average are not performed (The correction terms mentioned in last section is not considered in the real calculation.) because we only want to show the effectiveness of the subtraction methods in reducing the variance.  Since the computational time is directly proportional to the variance, we will use $V_{NONSUB}/V_{SUB}$ to repesent the performance for each subtraction method. The result of the local scalar is shown in figure\ref{scalar}.
We compare the results of polynomial subtraction method and perturbative method for $O(\kappa^{4})$, $O(\kappa^{7})$ and $O(\kappa^{10})$. The calculations are averaged across 10 gauge configurations ($\beta=6.0$), separated by 2000 sweeps.

\begin{figure}
 \begin{center}
  \includegraphics[scale=.75]{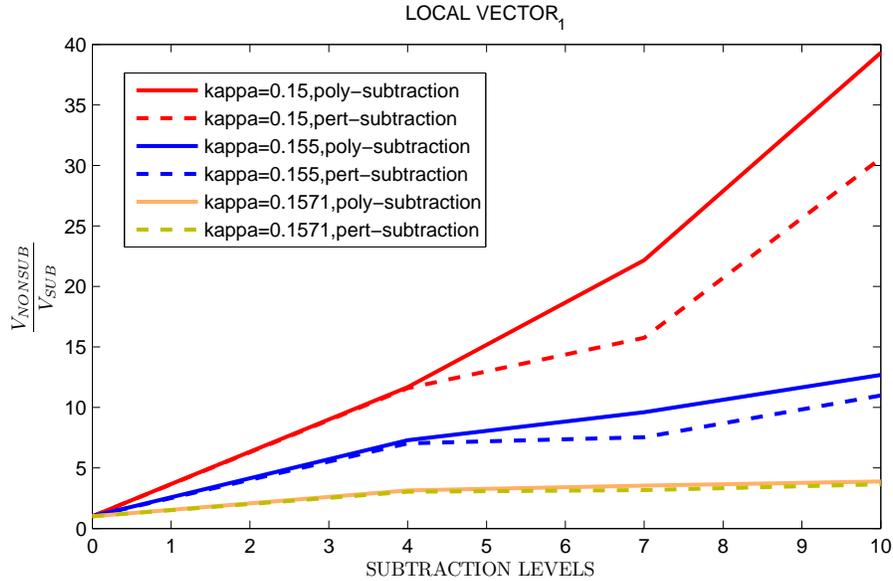}
 \end{center}
 \caption{Subtracted result for $Local\, Vector_{1}$.}
 \label{localJ1}
\end{figure}

\begin{figure}
 \begin{center}
  \includegraphics[scale=.75]{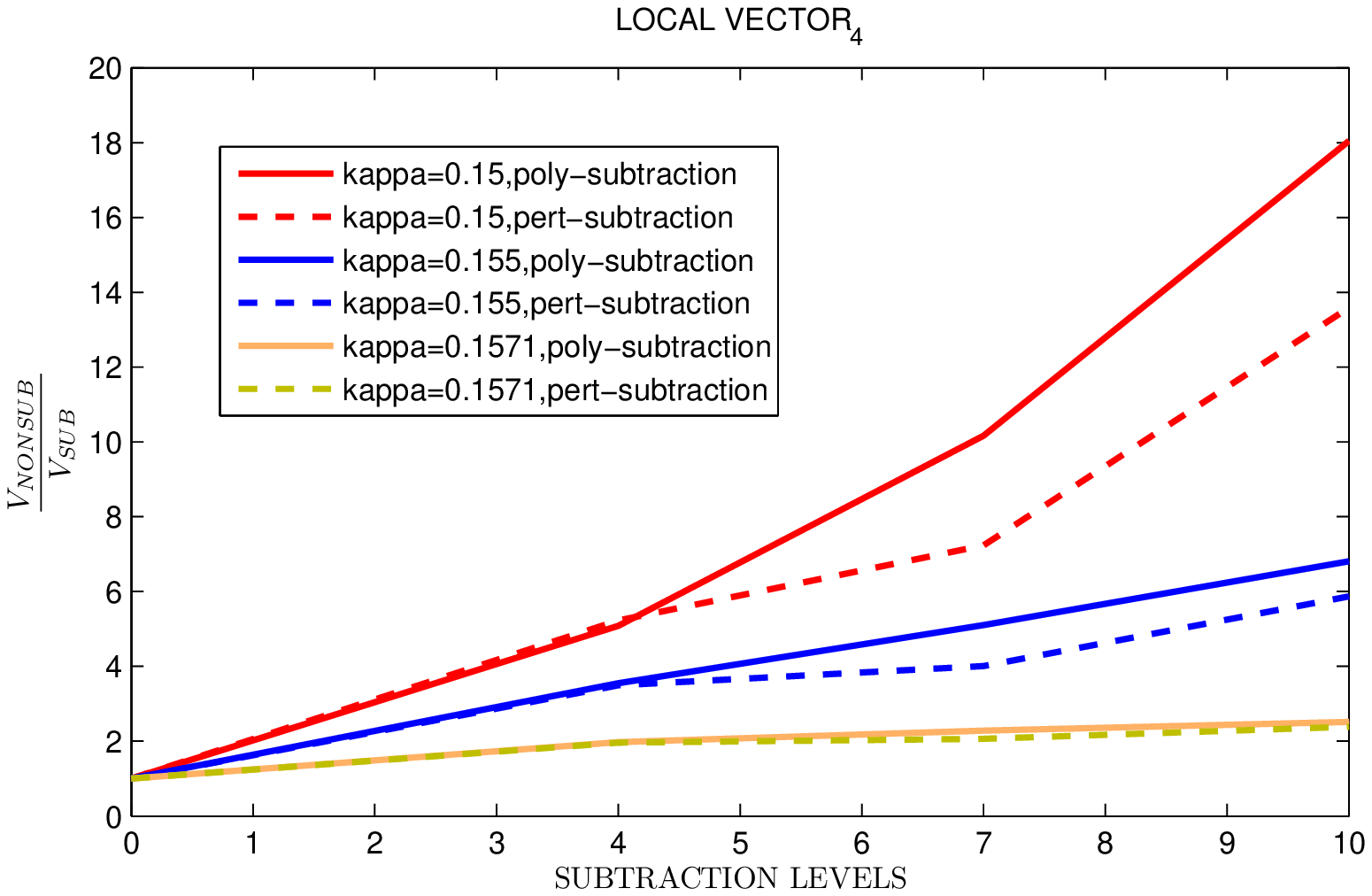}
 \end{center}
 \caption{Subtracted result for $Local\, Vector_{4}$.}
 \label{localJ4}
\end{figure}

\begin{figure}
 \begin{center}
  \includegraphics[scale=.75]{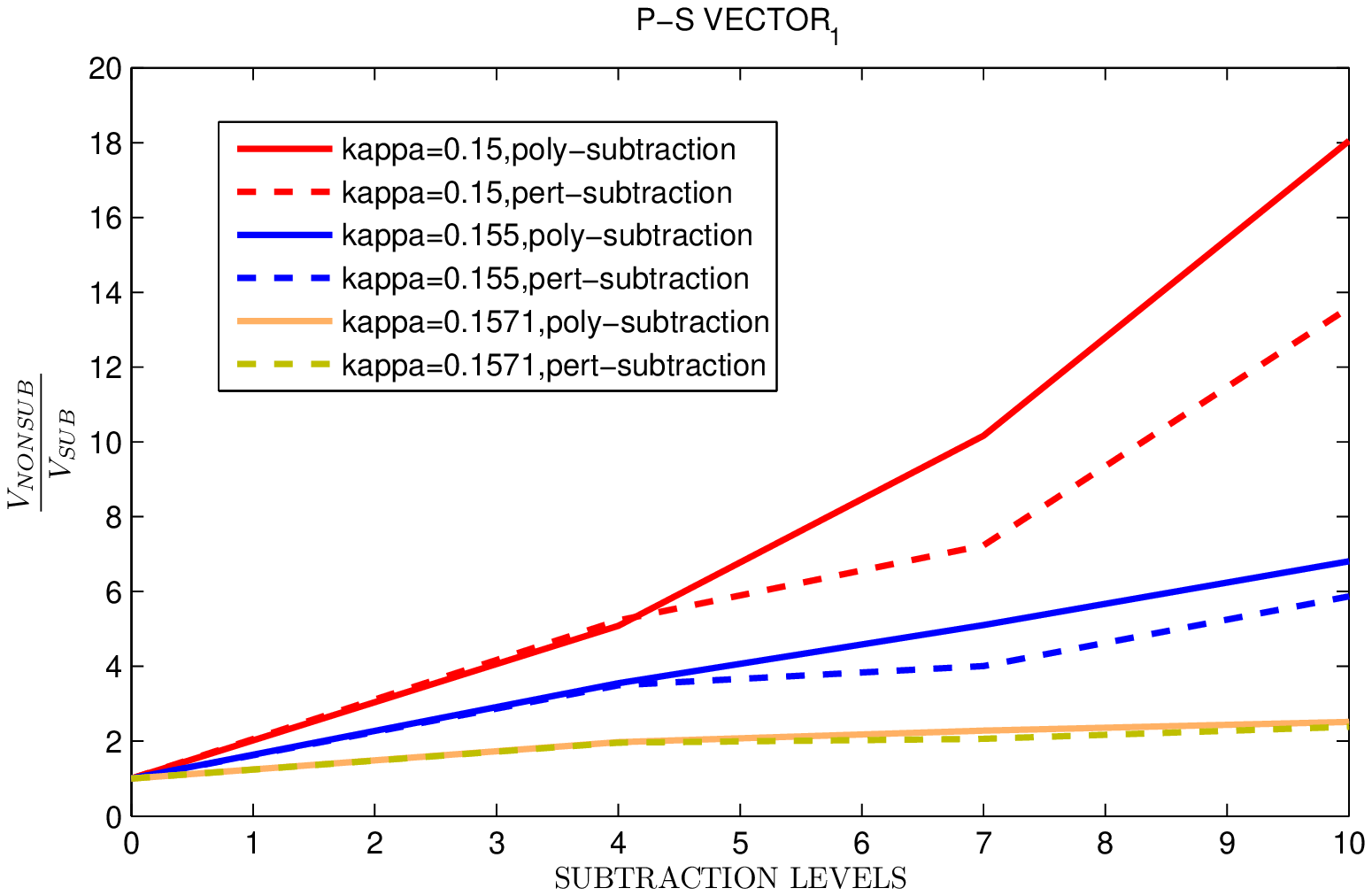}
 \end{center}
 \caption{Subtracted result for $Point-Split\, Vector_{1}$.}
 \label{ps1}
\end{figure}

\begin{figure}
 \begin{center}
  \includegraphics[scale=.75]{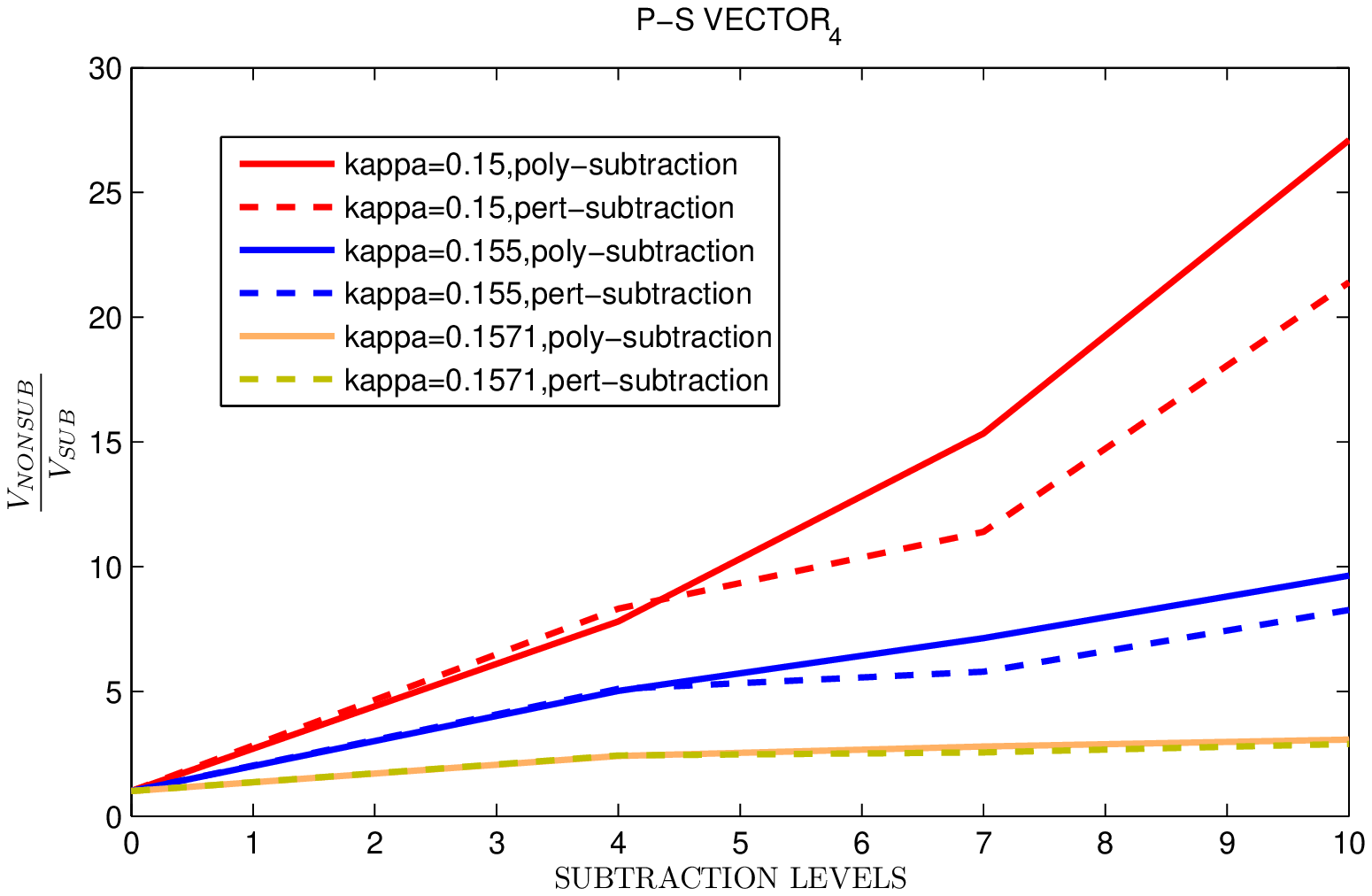}
 \end{center}
 \caption{Subtracted result for $Point-Split\, Vector_{4}$.}
 \label{ps4}
\end{figure}

The result shows that the polynomial subtraction outperforms the perturbative subtraction consistently for all the value of $\kappa$ at different orders. The ratio of  the variance is 9.5 for the polynomial subtraction vesus 7.4 for the perturbative subtraction at $10^{th}$ order of $\kappa$ with $\kappa=0.15$. When $\kappa$ is increased to 0.155, the benefit of polynomial subtraction is diminished. The variance ratio is 3.5 for polynomial subtraction and 3.1 for perturbative subtraction. At $\kappa_{critical}=0.1571$, the variance ratio for polynomial subtraction is 1.61 versus 1.52 for perturbative subtraction. The performance of the two methods are almost the same. The maximum \%reduction of variance is acheived at  $7^{th}$ order of $\kappa$ for each $\kappa$ value.
 The results of local vectors are shown from figure\ref{localJ1} to figure\ref{localJ4}. We can see the subtraction methods are more effective for local vector than scalar because the variance ratios are generally larger. Similarly, we can see the relative performance of the polynomial subtraction method is best for $\kappa=0.15$ and becomes less effective as $\kappa$ gets bigger. The results of point-split vectors are shown in figure\ref{ps1} to figure\ref{ps4}. The performance for  point-split operators is between local vector and local scalar.This can be seen from the variance ratio. And it's also more effective for small $\kappa$.
\begin{figure}
 \begin{center}
  \includegraphics[scale=.75]{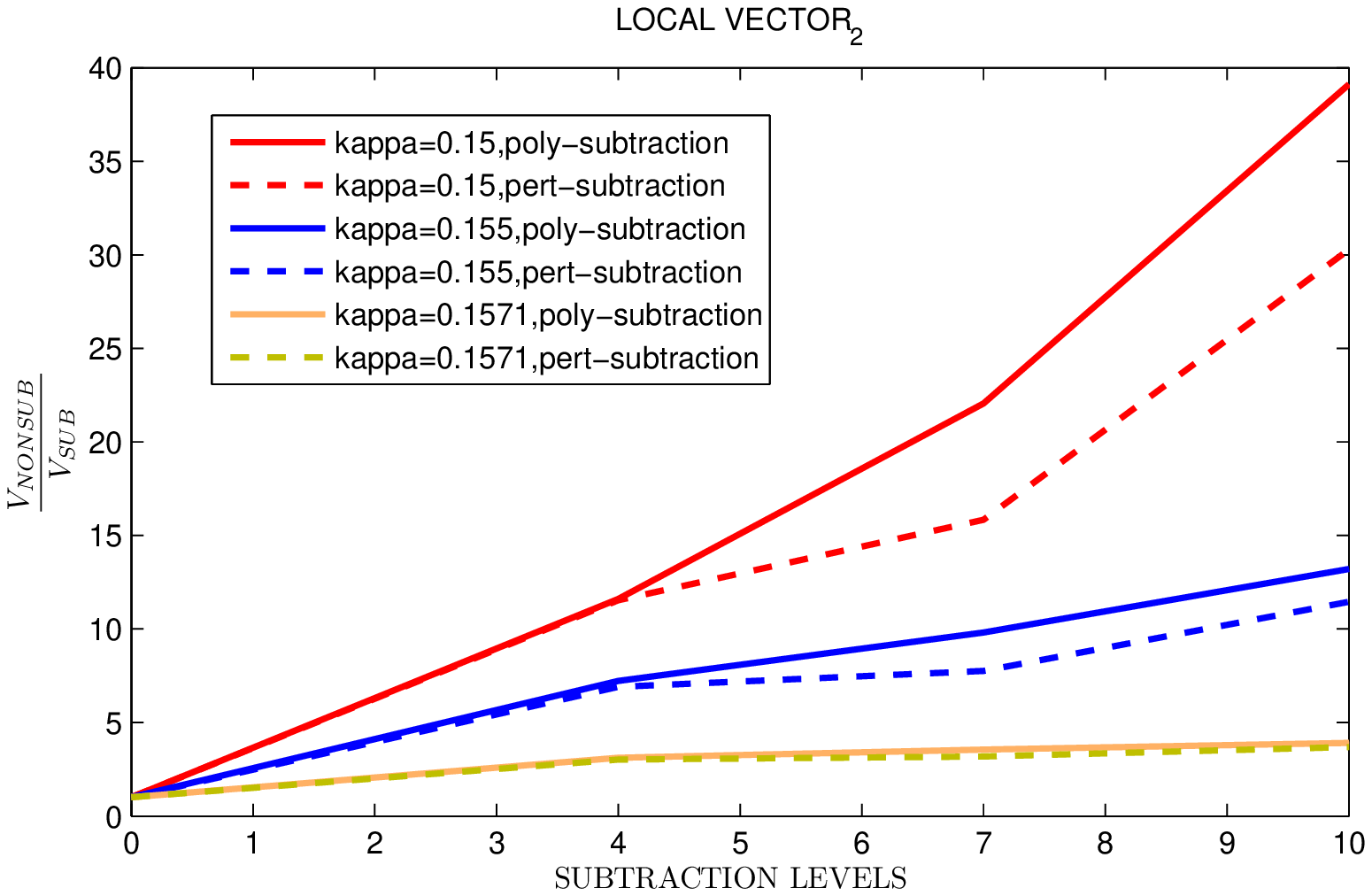}
 \end{center}
 \caption{Subtracted result for $Local \, Vector_{2}$.}
 \label{localJ2}
\end{figure}
\begin{figure}
 \begin{center}
  \includegraphics[scale=.75]{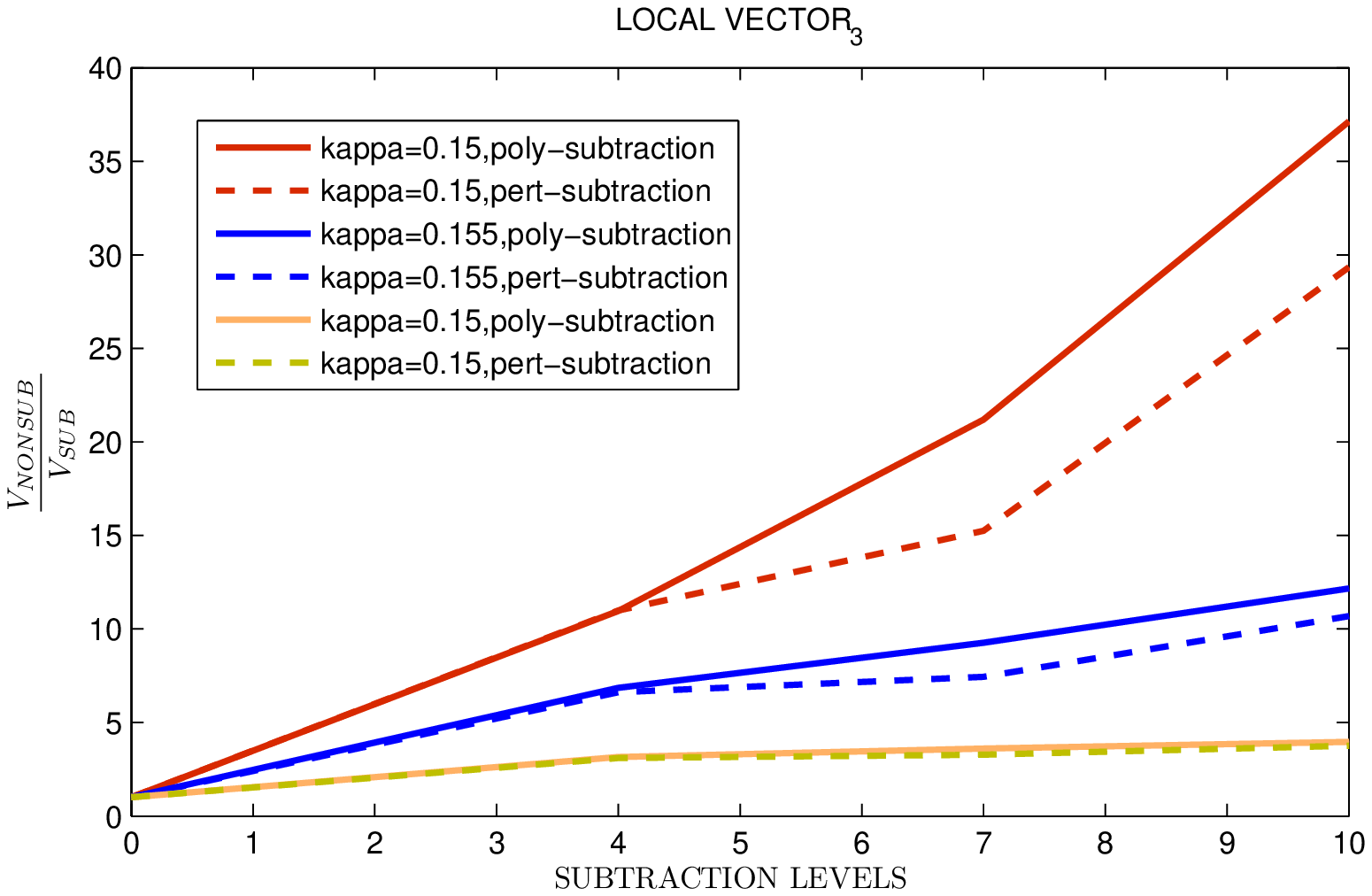}
 \end{center}
 \caption{Subtracted result for $Local\, Vector_{3}$.}
 \label{localJ3}
\end{figure}

\begin{figure}
 \begin{center}
  \includegraphics[scale=.75]{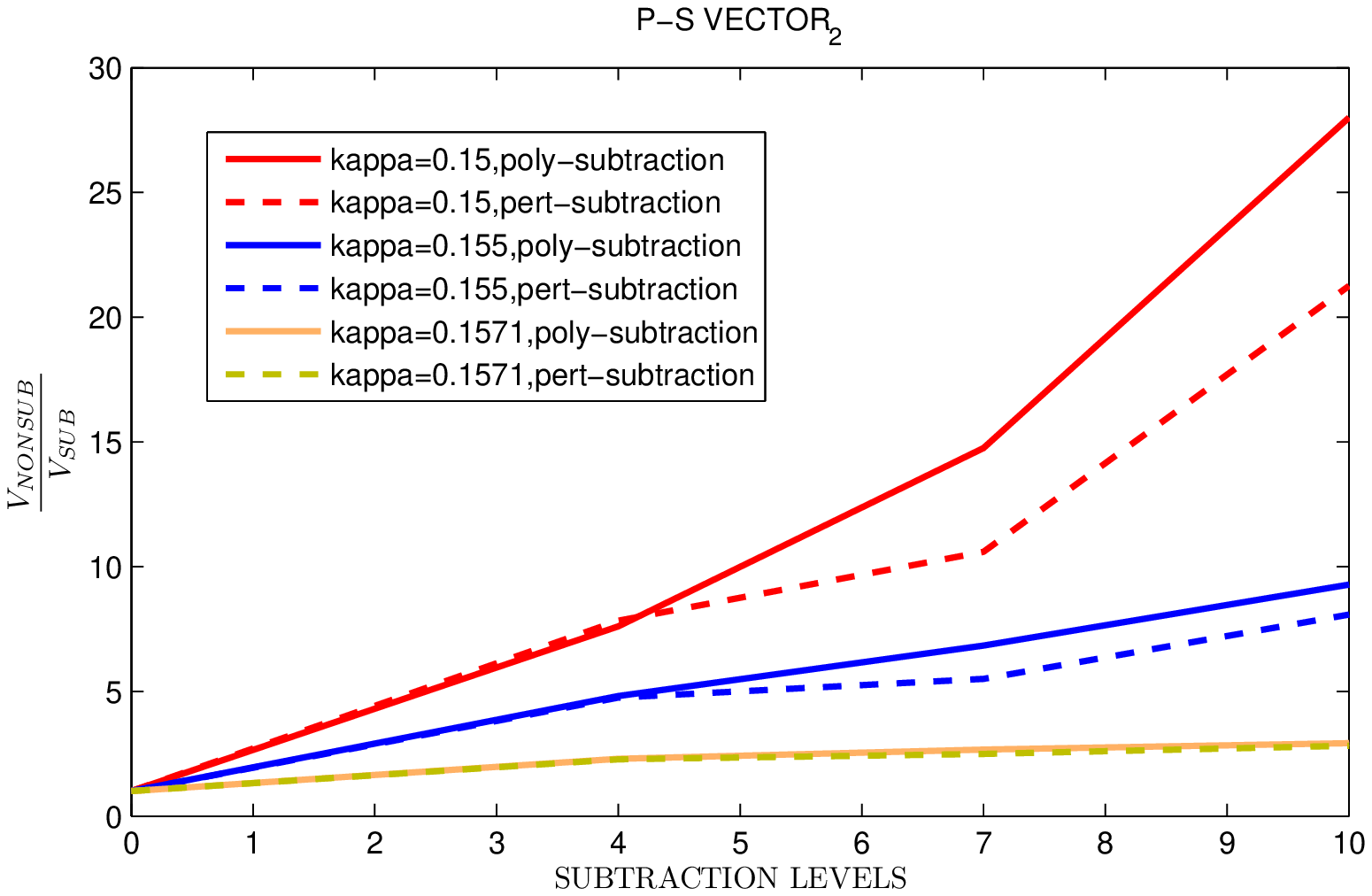}
 \end{center}
 \caption{Subtracted result for $Point-Split\, Vector_{2}$.}
 \label{ps2}
\end{figure}
\begin{figure}
 \begin{center}
  \includegraphics[scale=.75]{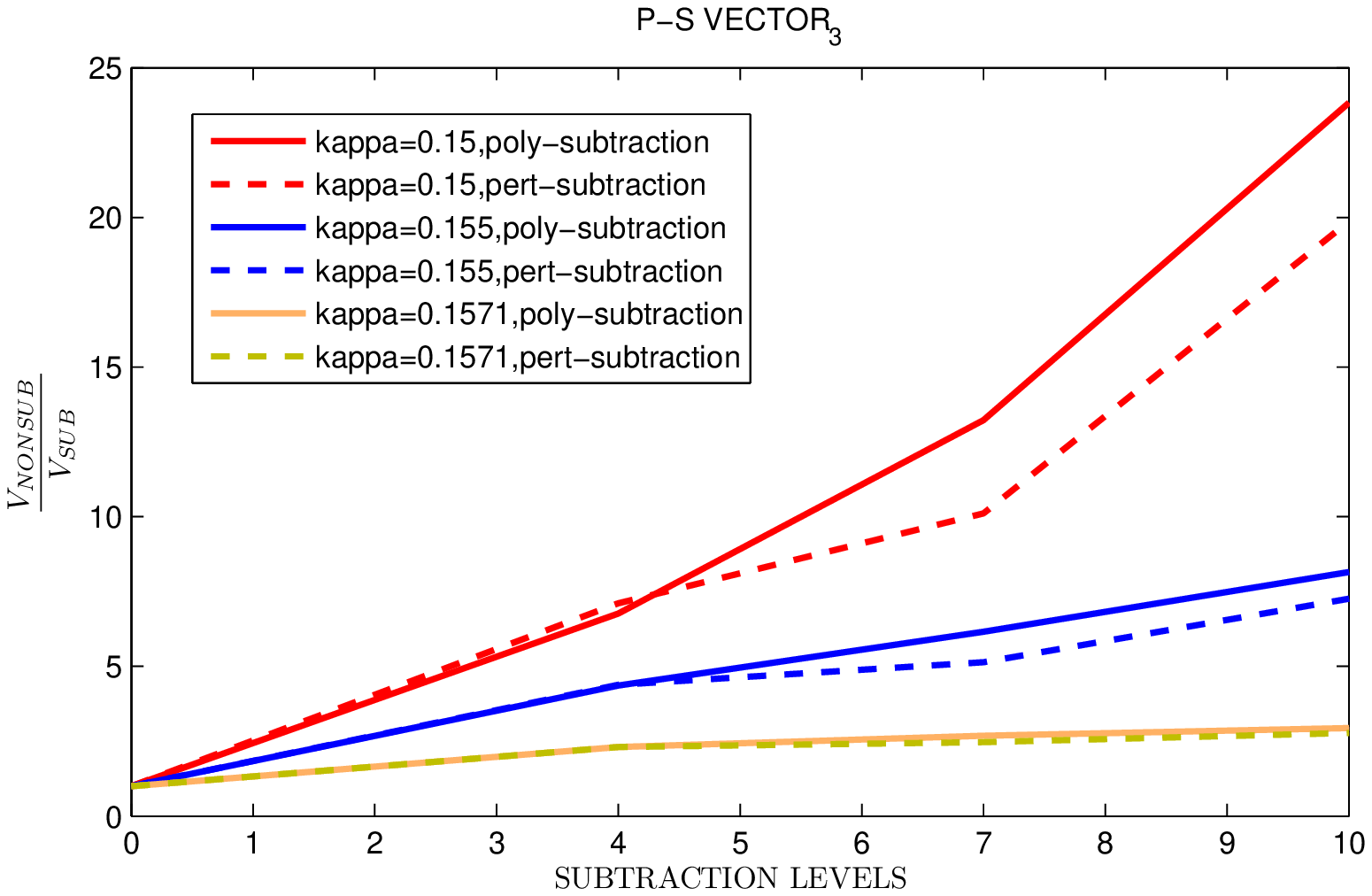}
 \end{center}
 \caption{Subtracted result for $Point-Split\, Vector_{3}$.}
 \label{ps3}
\end{figure}

The relative performance between the two methods can be measured by the ratio of $V_{PERT}/V_{POLY}$, where $V_{PERT}$ is the variance of perturbative subtraction and $V_{POLY}$ is the variance of polynomial subtraction. The relative ratio is summarized in table \ref{small} to table \ref{largeM}. We can see from all the three tables that the relative variance ratio is close to 1 at $4^{th}$ order of $\kappa$. This suggests the $4^{th}$ minimal residual polynomial is not a better approximation for $M^{-1}$ than the $4^{th}$ order perturbative expansion. This result is reasonable since we usually need to build a Krylov subspace with higher order of M to get larger reduction in the residual in Eq.(\ref{res}).

As we increase the $\kappa$, the variance ratio keeps decreasing. It means the polynomial subtraction suffers from the same problem as the perturbative subtraction. This is not surprising since both methods are expansion of $\kappa$s. The difference is the dependence of $\kappa$ is explicit for perturbative subtraction but implicit for polynomial subtraction.

\begin{table}
\caption{Relative performance for $\kappa=0.15$} 
\begin{center}
\begin{tabular}{|c|c|c|c|} \hline \hline
\multicolumn{4}{|c|}{$V_{PERT}/V_{POLY}$ at $\kappa=0.15$} \\ \hline \hline
Operators &$4^{th}\, Order \,of\, \kappa$ &$7^{th}\,Order\,of\,\kappa$ &$10^{th}\,Order\,of\,\kappa$ \\ \hline
Local Scalar &1.06 &1.40 &1.28 \\ \hline
Local Vector1 &1.01 &1.41 &1.29 \\ \hline
Local Vector2 &1.01 &1.39 &1.29 \\ \hline
Local Vector3 &1.00 &1.39 &1.27 \\ \hline
Local Vector4 &0.97 &1.41 &1.33 \\ \hline
Point-Split Vector1 &0.95 &1.37 &1.28 \\ \hline
Point-Split Vector2 &0.97 &1.39 &1.32 \\ \hline
Point-Split Vector3 &0.95 &1.31 &1.20 \\ \hline
Point-Split Vector4 &0.94 &1.34 &1.27 \\ \hline
\end{tabular}
\end{center}
\label{small}
\end{table}
\begin{table}
\caption{Relative performance for $\kappa=0.155$} 
\begin{center}
\begin{tabular}{|c|c|c|c|} \hline \hline
\multicolumn{4}{|c|}{$V_{PERT}/V_{POLY}$ at $\kappa=0.155$} \\ \hline \hline
Operators &$4^{th}\, Order \,of\, \kappa$ &$7^{th}\,Order\,of\,\kappa$ &$10^{th}\,Order\,of\,\kappa$ \\ \hline
Local Scalar &1.09 &1.26 &1.15 \\ \hline
Local Vector1 &1.04 &1.28 &1.16 \\ \hline
Local Vector2 &1.05 &1.26 &1.15 \\ \hline
Local Vector3 &1.03 &1.25 &1.14 \\ \hline
Local Vector4 &1.02 &1.27 &1.16 \\ \hline
Point-Split Vector1 &0.98 &1.24 &1.16 \\ \hline
Point-Split Vector2 &1.01 &1.24 &1.14 \\ \hline
Point-Split Vector3 &0.99 &1.20 &1.12 \\ \hline
Point-Split Vector4 &0.98 &1.23 &1.17 \\ \hline
\end{tabular}
\end{center}
\label{medium}
\end{table}
The best relative performance is at $7^{th}$ order of $\kappa$ for each $\kappa$. The relative reduction of variance is quite consistent across different operators at same level of subtraction. For example, at $\kappa$=0.15, all the variance ratios for the $7^{th}$ order subtraction are within 1.30$\sim$1.40. The performance of local scalar and local vector vectors is slightly better than that of the point-split vectors but the difference is not statistically significant. It is unknown why the $10^{th}$ order subtraction underperforms the $7^{th}$ order subtraction. There could be some numerical saturation at the $7^{th}$ order subtraction. This could be further investigated by calculating more orders' of subtraction and analysing the relative performance. Take local scalar as an example. The variance ratio of 1.40 at $\kappa$=0.15 for the $7^{th}$ order subtraction means a 29\% relative reduction in variance. Even though the ratio is reduced to 1.10 at $\kappa$=0.1571, we still have a 9\%  relative reduction in variance. This demonstrates the effectiveness of polynomial subtraction method in reducing the variance of calculation.

\begin{table}
\caption{Relative performance for $\kappa=0.1571$} 
\begin{center}
\begin{tabular}{|c|c|c|c|} \hline \hline
\multicolumn{4}{|c|}{$V_{PERT}/V_{POLY}$ at $\kappa=0.1571$} \\ \hline \hline
Operators &$4^{th}\, Order \,of\, \kappa$ &$7^{th}\,Order\,of\,\kappa$ &$10^{th}\,Order\,of\,\kappa$ \\ \hline
Local Scalar &1.04 &1.10 &1.06 \\ \hline
Local Vector1 &1.04 &1.12 &1.06 \\ \hline
Local Vector2 &1.03 &1.12 &1.06 \\ \hline
Local Vector3 &1.02 &1.10 &1.05 \\ \hline
Local Vector4 &1.01 &1.10 &1.05 \\ \hline
Point-Split Vector1 &1.01 &1.09 &1.06 \\ \hline
Point-Split Vector2 &1.01 &1.07 &1.03 \\ \hline
Point-Split Vector3 &1.00 &1.09 &1.05 \\ \hline
Point-Split Vector4 &1.01 &1.10 &1.05 \\ \hline
\end{tabular}
\end{center}
\label{largeM}
\end{table}
\begin{table}
\caption{Relative performance for large matrix} 
\begin{center}
\begin{tabular}{|c|c|c|c|} \hline \hline
\multicolumn{4}{|c|}{$V_{PERT}/V_{POLY}$ for $24^{3}\times 32$ Lattice} \\ \hline \hline
Operators &$\kappa=0.15$ &$\kappa=0.155$ &$\kappa=0.1571$ \\ \hline
Local Scalar &1.39 &1.28 &1.08 \\ \hline
Local Vector1 &1.37 &1.25 &1.08 \\ \hline
Local Vector2 &1.39 &1.25 &1.08 \\ \hline
Local Vector3 &1.39 &1.25 &1.06 \\ \hline
Local Vector4 &1.39 &1.28 &1.08 \\ \hline
Point-Split Vector1 &1.37 &1.23 &1.08 \\ \hline
Point-Split Vector2 &1.37 &1.25 &1.08 \\ \hline
Point-Split Vector3 &1.39 &1.25 &1.08 \\ \hline
Point-Split Vector4 &1.37 &1.25 &1.08 \\ \hline
\end{tabular}
\end{center}
\label{large}
\end{table}
Based on the analysis for the $16^{4}$ lattice, we implement the polynomial subtraction on a $24^{3}\times 32$ lattice, which corresponds to a $2.6$ million $\times$ $2.6$ million matrix. We only perform the calculation for the $7^{th}$ order subtraction since it is the most effective subtraction level. The result is shown in table \ref{large}. We can see that these results quite agree with what we have found in the small lattice case. The variance ratios for each operator are around 1.4, 1.3, 1.1 for $\kappa$=0.15, 0.155, 0.1571 respectively, which means it's more effective for smaller $\kappa$.

\section{Conclusion}\label{Ssix}
From the testing results on the $16^{4}$ and $24^{3}\times 32$ matrices, we can conclude that the polynomial subtraction can save the computational time by about 30\% compared to the perturbative subtraction method for small $\kappa$. The benefit is reduced to about 10\% at $\kappa_{critical}$. Although the  polynomial subtraction suffers from the same problem as the perturbative subtraction due to the dependence of $\kappa$, it's still competitive since it doesn't require much extra calculation time to get the coefficients but can reduce the variance or computational time significantly at least for small $\kappa$. The techniques of combining the perturbative subtraction and eigenspectrum subtraction\cite{vicpaper},which is developed to solve the problem for largre $\kappa$, has been successfully tested for small QCD lattice($8^{4}$) on matlab by our group. Since polynomial subtraction is more robust than perturbative subtraction, it is intuitive to combine the polynomial and the eigenspectrum subtraction to get a better result. This could be our next work in the future.
\newpage
\bibliography{referfile}{}
\bibliographystyle{apsrev}

\end{document}